# Plasmonic and silicon spherical nanoparticle anti-reflective coatings

K.V. Baryshnikova,[1] M.I. Petrov,[1,2] V.E. Babicheva,[1,3] and P.A. Belov[1]
[1]ITMO University, Kronverkskiy, 49, St. Petersburg 197101, Russia
[2]St. Petersburg Academic University, Russia
[3] Center for Nano-Optics, Georgia State University, P.O. Box 3965, Atlanta, GA 30302

Over the last decade, plasmonic antireflecting nanostructures have been extensively studied to be utilized in various optical and optoelectronic systems such as lenses, solar cells, photodetectors, and others. The growing interest to all-dielectric photonics as an alternative optical technology along with plasmonics motivates us to compare antireflection properties of all-dielectric and plasmonic nanoparticle coatings based on silver and crystalline silicon. Our results of numerical simulations for periodic arrays of spherical nanoparticles on top of amorphous silicon show that both silicon and silver nanoparticle coatings demonstrate strong anti-reflective properties in the visible spectral range. In this work, we show for the first time that blooming effect, that is zero reflection from the structure, with silicon coatings originates from the interference of electric- and magnetic-dipole responses of nanoparticles with the wave reflected from the substrate, and we refer to it as substrate-mediated Kerker effect. For the silver coating, our results agree with previously observed substrate-induced bi-anisotropy and blooming, caused by substrate-induced magnetic response. Finally, we numerically show high effectiveness of silicon and silver coatings for the application in thin-film photovoltaic elements, which is related to the suppression of reflection from the high-index substrate and increased light absorbance in the active layer with coating. Depending on the nanoparticle size, either silicon or silver coating is more efficient, and overall increase of absorption up to 30% can be achieved.

## Introduction

The suppressing of light reflectance from a flat surface has been an important technological problem for the last decades. The methods of canceling reflection rely on different elements such as simple quarter-wavelength dielectric layers, nanostructured surfaces for light trapping, graded-index layers, and others [1]. Various advantages of plasmonic nanostructures were found out recently [2], and most of them are related to excitation of the intense localized surface plasmon resonance (LSPR) in metallic nanostructures and strong suppression of light reflectance in vicinity of the resonance [3]. Despite the active studies on these topics and great promises of the applications, practical use of the plasmonic nanoparticles is still hindered by many challenges, for instance one need to reduce large ohmic losses of plasmonic materials, which suppress nanoparticle resonances, or overcome surface oxidizing, which changes the optical properties of nanostructures [4]. Moreover, LSPR depends on the shape of the specific nanoparticle rather than its size and inter-particle distance, which makes the tuning of optical properties problematic [5]. Since recently the all-dielectric photonics has been suggested as an effective alternative to plasmonics [1][6][7]. The concept is based on high-refractive index nanostructures, which possess magnetic Mie resonance along with electric one and allow a simultaneous control of magnetic and electric components of light on the nanoscale [8]. Silicon is considered as a promising material for all-dielectric photonics having one of the highest possible refractive index in the visible and near-infrared wavelength ranges and relatively small optical losses at the same time [9]. Resonance spectrum of high-index structures is defined by geometrical parameters along with optical properties and consequently can be efficiently tuned during the fabrication process [10]. Furthermore, researchers' attention has recently been attracted by important features of high-index nanoparticles: at a certain wavelength they possess a high directionality of radiation pattern [11][12], which results in strong forward and low back scattering. This behavior was predicted for particles with equal magnetic and electric dipole moments by Kerker and colleagues [13] while such particles are often referred as Huygens element [14][15][16]. Besides possible applications in photovoltaics, Huygens elements are suggested as main element of metasurfaces and future flat photonic devices for efficient light manipulations on the nanoscale [17].

In this manuscript, we are aiming on comparison of LSPR and Mie-resonant dielectric coatings of high-index optical elements, e.g. silicon-air interface, based on nanoparticles of silver and silicon as the most favorable plasmonic and all-dielectric photonics materials, respectively. We clarify the physical background of anti-reflective properties of such nanoparticle arrays, and more importantly, for both plasmonic and dielectric nanoparticle coatings, we show the key role of high reflection from the layer beneath array to suppress overall reflection from the coated structure. In particular, we demonstrate that the antireflection effect of silicon coatings originates from destructive interference of wave reflected from the substrate with the fields reradiated by the electric and magnetic dipoles induced in silicon nanoparticles. This effect can be considered as substrate-mediated Kerker-effect as an analogue of well-studied Kerker effect for dielectric nanoparticles in vacuum or homogeneous environment (low-index substrates and matched-index covering of nanoparticle array). Finally, we show that light harvesting can be improved by placing silicon nanoparticles coatings on top of silicon solar cell or photodetector. We calculate change of absorption inside photovoltaic elements in both broad and narrow bands and discuss efficiency of the coating.

## Results
### 1. Structure model and simulation parameters

According to the Mie theory, single silicon particles possess dielectric-type dipole resonances at the wavelength $\lambda$ fulfilling the condition $\lambda \approx 2Rn$, where $R$ is the nanoparticle radius and $n$ is the real part of the silicon refractive index [9]. Silicon particles with radius 40-80 nm have only electric (ED) and magnetic dipole (MD) resonances in the visible spectral range, and the higher multipole modes are in the ultra-violet range being weak and broad. Silver particles support LSPR of the dipole mode, which is defined mainly by the permittivities of metal $\varepsilon_\text{m}$ and surrounding material $\varepsilon_\text{sur}$, and particularly, for dipole LSPR in spherical nanoparticle permittivities should satisfy the condition $\varepsilon_\text{m} + 2\varepsilon_\text{sur} = 0$, which in case of silver occurs in the ultra-violet range.

We consider a square lattice of crystalline silicon (c-Si) and silver spherical nanoparticles (see Fig.1) placed on the surface of hydrogenated amorphous silicon substrate (a-Si:H), which is widely-used material in thin-film solar cells and photodetectors [18]. To compare the efficiency of crystalline silicon and silver nanoparticle arrays for transmission enhancement, we restrict ourselves to the case of normal incidence of light and consider the particles of 15-80 nm radius. In this paper, all results are obtained for the fixed value of surface filling factor $f = \pi R^2 / d^2 = 0.18$ making possible comparing structures with different nanoparticles size and keeping the dipole-dipole interaction of neighboring nanoparticles rather weak and constant for each coating material. Numerical simulations and postprocessing were performed in CST Microwave Studio and Comsol Multiphysics software packages. The permittivities of silver and silicon materials were taken from the experimental measurements [18][19][20].

## 2. Substrate-mediated zero backscattering from nanoparticle array

In order to identify the nanoparticle resonances, we have calculated the absorbance inside the nanoparticles. The particular set of parameters: for Si-NPC $R_{Si}$ = 60 nm and $d_{Si}$ = 250 nm, and for Ag-NPC $R_{Ag}$ = 30 nm and $d_{Ag}$ = 125 nm, provides the condition for the excitation of all resonances in the range of silicon photovoltaics operation ($\lambda$ < 600 nm). The absorbance inside the nanoparticles shows spectral position of nanoparticle resonant modes. Figure 2(a) shows the LSPR for metal nanoparticles (labeled "ED") and two Mie-resonances for silicon nanoparticles – electric (at shorter wavelength, labeled "ED") and magnetic (at longer wavelength, labeled "MD"). For the chosen radiuses, absorption resonances of silicon and silver nanoparticles have similar width and comparable amplitude. Spectral position of the LSPR resonance in silver nanoparticles is in a good agreement with resonance condition for spherical nanoparticle in vacuum ($\mathrm{Re}(\varepsilon_m) \approx -2$), and the slight splitting of the peak results from the interaction of the nanoparticle with the high-index substrate and induced magnetic response which we discuss below in more detail.

Now, let us analyze reflection and transmission spectra of nanoparticle arrays and influence of high-index substrate on the blooming effect. The calculated reflectance spectra of Ag-NPC and Si-NPC demonstrate strong antireflectance properties, which manifest itself in sharp decreases of reflectance at the wavelength 400 nm for Ag-NPC and 480 nm for Si-NPC (Fig. 2(c)). The interplay of magnetic and electric dipole resonant modes in dielectric nanostructures was experimentally observed, for example, through the asymmetry of light scattering by individual silicon nanoparticles [11]. This feature is explained by Kerker effect [12][21], which was predicted for optical elements with electric and magnetic dipole moments equal both in amplitudes and in phase. Such elements are often referred as Huygens elements [12][17], as equality of electric and magnetic moments means that effective impedance of the metasurface (proportional to $\sqrt{(\mu/\varepsilon)}$) is equal to the vacuum impedance, which ensures transmittance through the array of elements in vacuum without reflection.

The strong unidirectional scattering of Huygens elements, and consequently its drastic influence on the transmission spectrum of the whole array, provides the basis for engineering metasurfaces for flat nanophotonic devices. For instance, in the recent works, silicon nanodisks were studied in homogeneous low-index environment [17][22], and the possibility of light phase change by simultaneous controlling of magnetic and electric dipole resonances was demonstrated. In fact, the vast majority of studies were performed for nanoparticles on top of low-index substrate or theoretically considering them in vacuum. The detailed studies [23] revealed that the Kerker effect, which manifest itself in high forward-to-backward ratio of scattering from single nanoparticles, is significantly modified and partially suppressed by placing the nanoparticle over the substrate with high index. Moreover, in that case, dipole moment modes of the studied nanocylinders suffer from high leakage to the substrate, which causes a decrease of total scattering.

To establish the role of magnetic and electric dipoles' scattering in the antireflection effect for the case of nanoparticle array on silicon substrate, we have performed additional numerical simulations in order to separate the contributions of the NPC and the substrate in the total reflectance. First, we have calculated the reflectance spectrum for a plane wave normally incident at the array of silicon nanoparticles in air (shown in Fig. 3(a) with dashed line). The zero-reflection at point A at the long-wavelength side of the magnetic resonance corresponds to Kerker effect [11]. The electric and magnetic dipole moments are in-phase with each other, and equal in amplitude, which results in zero backscattering from the individual nanoparticles (Poynting vector $\mathbf{S}_{scat}$ points down, see Fig. 3(b)). On the contrary, between the ED and MD resonances, at the point B, one can see strong reflectance, which corresponds to strong backscattering because of almost $\pi$-shift between the electric and magnetic dipole moments (Poynting vector $\mathbf{S}_{scat}$ points up, see Fig. 3(c)). For the wavelength range between ED and MD resonances, the wave reflected from the nanoparticle array is shifted has $\pi/2$ phase shift relatively to the incident wave, but has opposite direction. Second, we calculated the wave reflected from a bare silicon substrate, and we have sum it up with the wave reflected from nanoparticle array in air. The wave reflected from the bare silicon substrate has $\pi$-phase shift relatively to the incident wave and additional $2kR \approx \pi/2$ phase incursion, which after summation with the wave scattered by the nanoparticle array cancel each other and give zero reflectance between ED and MD resonances: the resulting curve is shown in Fig. 3(a) with solid line. Comparing the obtained reflectance spectrum with the spectrum calculated from the full simulation of the Si-NPC over the silicon substrate, one can see a good quantitative agreement with our simple model.

To sum up, our calculations for the nanoparticle array on Si substrate show that there exists a wavelength range with drastic decrease of reflectance (430-480 nm), which results from the destructive interference of magnetic and electric dipole scattered waves and the wave reflected from the substrate. In this sense, the predicted antireflection can be referred as substrate mediated Kerker effect.

Further, let us clarify origin of blooming effect in the array of plasmonic nanoparticles (see Fig. 2(c)), which have only electric dipole resonance. The antireflectance effect for small metal nanoparticle array has been recently studied and associated with the substrate-induced bi-anisotropy [24][25]. The electric dipole moment of the nanoparticle induces charge distribution in the high-index silicon substrate in the vicinity of nanoparticle in the so-called "hot spots". Though the single metal particle does not possess any intrinsic magnetic dipole moment, the overall charge distribution inside the particle and in the substrate hot-spot has non-zero magnetic moment. In this case, the destructive interference of the field scattered by ED and induced MD moments the wave reflected from the silicon substrate causes blooming effect. Induced MD moment causes increase of absorption, which can be either spectrally overlapped with ED peak or slightly separated depending on how high refractive index of the substrate. In our case of the silicon substrate, we observe a negligible splitting in the absorption spectra in Fig. 2(a). One can mentioned that good agreement between the simulation results and the simple interference model for the case of Si-NPC (Fig. 3a) also shows that in contrast to the case of Ag-NPC, the interaction of Si-NPC with the substrate is rather weak and no pronounced bi-anisotropy effect is present in the case of silicon nanoparticles over the high-index substrate.

To sum up, the origin of antireflection for the case of high-index substrate is similar to well-known Kerker effect. For silicon particles embedded in homogeneous optical environment, Kerker effect either occurs on the long-wavelength side of the MD resonance or on small-wavelength side of ED [11], and in both cases, it is sensitive to indexes of substrate and superstrate of the nanoparticle. In contrast, high-index substrate, which is the case of the present work, possesses sufficient reflection, and the substrate-mediated Kerker-like effect manifests itself in the destructive interference of the ED and MD fields with the wave reflected from the substrate surface, and the blooming is observed between the ED and MD resonances.

## 3. Transmittance efficiency of plasmonic and silicon nanoparticle coatings

The antireflective properties discussed above can be applied to increase absorption inside photovoltaic elements. Further, we calculate efficiency of NPC in terms of increased transmission and show that both types of coating can provide better light trapping. The transmittance to the substrate is shown in Fig. 2(b) and is calculated right beneath the nanoparticle array as shown in the inset. Comparing Fig. 2(a) and 2(b), one can see that for metal nanoparticles, transmittance to the substrate decreases at the LSPR resonance due to the strong absorption inside nanoparticles (reported earlier in *e.g.* [26]). The same effect is observed for Si-NPC, where the absorption inside nanoparticles is resonantly increased at the wavelengths of ED and MD resonances. Thus, it is important to stress that ED or MD resonance itself does not bring increase of transmittance and this wavelength cannot be used to increase efficiency of light harvesting. However, both silver-nanoparticle coating (Ag-NPC) and silicon one (Si-NPC) enhance transmission to the substrate at the wavelengths longer than ED and MD resonances, respectively (Fig. 2(b)). It originates from the effect of constructive interference of scattered light with incident wave. Indeed, for wavelengths longer than the resonant, dipole oscillations occur in-phase with incident light, while below the resonant wavelength, dipole oscillates with $\pi$-shift in respect to the driving field, which results in destructive interference observed for shorter wavelengths $\lambda \approx 300\text{-}350$ nm for Si-NPC [see Fig. 2(b)].

To compare the antireflective efficiency of different nanoparticle coatings, we introduce integral transmission enhancement as follows:

$$F_I = \left( \frac{\int T(\lambda) d\lambda}{\int T_0(\lambda) d\lambda} - 1 \right) \cdot 100\%, \tag{1}$$

where $T(\lambda)$ and $T_0(\lambda)$ are the transmittances calculated with and without coating, respectively; spectral integration is taken over the wavelength range from 300 to 800 nm.

The dependence of the integral enhancement on the nanoparticle radius is plotted for different types of coatings in the Fig. 4(b). We see the non-monotonous dependence of the integral transmission enhancement for Si-NPC: the smaller nanoparticles show higher efficiency (up to 8%) than the larger ones, which occurs due to the stronger absorption inside the nanoparticles. The quadrupole response of the silicon nanoparticles with the sizes considered in this work is rather weak and broad because of the high loss of silicon in ultra-violet range. For Ag-NPC, although particles with radius $R < 20$ nm demonstrate negative transmission efficiency $F_I$, we see the increase of the efficiency with the increase of nanoparticle sizes that is 8% for particles with $R = 50\text{-}60$ nm and reaches 15% for $R = 80$ nm. Here we should note that silver particles with the radius more than 30-40 nm demonstrate strong higher-order multipole response [26]. Thus, plasmonic and silicon nanoparticles of the same size demonstrate different optical properties: silver nanoparticles with large size allow multipolar-mode resonances, which increase integral transmission efficiency of the coating.

From practical point of view, silicon nanoparticles provide more advantages in comparison with silver ones. In particular, the experimentally measured efficiency of silver nanoparticle coatings are typically lower than predicted by Fig. 4 (b), which can be explained by severe degradation of silver in ambient conditions [4][27]. In contrast to silver, silicon has high chemical stability and lower optical losses, which makes them promising candidates for future nanophotonic devices. Moreover, recent advances in methods of their cost-effective fabrication by laser printing open up a possibility to vary nanoparticle size and crystallographic phase [28][29], making silicon nanostructures even more prospective for photovoltaic applications.

The suppression of the reflectance results in the resonant increase of the transmittance with Si-NPC, and we define $\lambda_p$ as the wavelength of maximum transmittance, so it is a function of nanoparticle radius $R$ [see Fig. 2 (b)]. The dipole resonances of silicon nanoparticles are defined by their sizes in accordance with Mie theory, and the transmission maximum $\lambda_p$ shifts toward long wavelength region with increasing the nanoparticles sizes [Fig. 4(a)]. One can notice that the position of $\lambda_p$ shifts along with the positions of ED and MD resonances staying located between them. For larger nanoparticles, resonances are stronger, and the increase of transmission is higher than for small nanoparticles. Thus, we would also like to pay attention to the narrow-band increase of transmittance and introduce the peak transmittance enhancement of Si-NPC as the enhancement at $\lambda_p$:

$$F_p = \left( \frac{T(\lambda_p)}{T_0(\lambda_p)} - 1 \right) \cdot 100\%, \tag{2}$$

which also can be plotted as a function of nanoparticle radius (see Fig. 4(b) secondary y-axis). One can notice that both ED and MD resonances are vanished for nanoparticles smaller than 40 nm radius, and resonant enhancement is not observed (see Fig. 4(a)). The value of the resonant enhancement changes not monotonically and has maximum $F_p = 41\%$ for $R = 65$ nm (see Fig. 4(b)). Although the resonant transmission at $\lambda_p$ increases with the increase of nanoparticle sizes, it also shifts to the longer wavelength region (λ > 500 nm). There, the transmission of the bare silicon substrate is higher, and thus, for nanoparticles with larger radiuses (R > 70 nm), the resulting transmission enhancement is decreased.

Overall, we observe that both narrow- and broad-band transmittance increases can be obtained with nanoparticle coatings. For the increase in the entire visible range, Ag-NPC with radiuses 70-80 nm give the best transmittance enhancement up to 15%, and Si-NPC with radiuses 30-40 nm causes up to 8% of enhancement. At the same time, at the narrow spectral range, Si-NPC enables up to 40% of transmittance increase.

### 4. Antireflective coating for solar cells

High cost of energy production is one of the main problems that prevent wide implementation of solar cells [30]. For conventional planar photovoltaic structures, the thickness of crystalline silicon layer varies from 160 to 240 μm [31]. Today, price of materials gives a big contribution to the net costs, and decreasing the thickness of the active layer can bring a significant benefit in lowering costs. For these purposes, thin-film technology that uses amorphous Si layers with thickness down to 150 nm is very advantageous [32]. Moreover, the thin-film technology has a number of advantages such as flexibility of the solar cells, possibility of covering functional surfaces, and other. However, comparing to conventional solar cells, thin films possess a lower capability of overall light-to-charge conversion. To improve it and increase the light absorption in the active layer, we suggest to apply the Si-NPC, and we compare their efficiency to silver plasmonic coatings [3][33].

The typical thin film solar cell contains the active p-i-n layers for charge separating and the substrate layer of transparent conductive oxide (in our case indium tin oxide, ITO) (Fig. 5). For the calculations, optical parameters of the silicon were chosen with the account on different doping levels [18], and optical parameters of ITO are taken from [34]. The absorbance in the active layer can be estimated as

$$\eta(\lambda) = \frac{1}{P_0} \int_{\substack{active \\ layer}} a(\lambda) |E|^2 d^3r \qquad (3)$$

where $a(\lambda)$ is the absorption coefficient of the active region, $E$ is the electric field intensity inside the active region, and $P_0$ is the incident flux density.

The results of the modeling with the proposed coatings are presented in the Fig. 6 (a,b). The spectra of nanoparticles absorption on top of the active layer [Fig. 6(a)] resemble the spectra of nanocoatings placed over the homogeneous amorphous silicon (Fig. 6(a), lines with circles) demonstrating pronounced electric and magnetic resonances. The calculated curves slightly differ from the absorption spectra shown in Fig. 2 as the optical constants of doped layers differ from undoped silicon.

The corresponding spectra of absorption in the active layer are shown in Fig. 6(b). The spectrum of absorbance inside the bare active layer is defined by the intrinsic losses in silicon, which decrease for wavelengths longer than 600 nm. The absorption spectra of the active layer with NPCs fully agree with spectra of light transmittance to the bulk silicon [see Fig. 2 (b)]. The light absorption in the active layer at the ED and MD resonances is suppressed due to strong absorption inside the nanoparticles, whereas at the wavelength of Kerker-type resonance we see a strong enhancement of light absorption. In the case of Si-NPC, the active layer absorption possesses narrow spectral character, which reproduces the spectral behavior of enhanced transmission similar to Fig. 4.

Similar to the case of homogeneous silicon substrate (Eq. (1)), in order to compare the efficiency of different coatings we have introduced a parameter of integral absorbance enhancement in the active region, which is directly related to generated carrier current: $G_I = \left(\int \eta(\lambda)d\lambda / \int \eta_0(\lambda)d\lambda - 1\right) \cdot 100\%$, where $\eta$ and $\eta_0$ are the total absorbance in the active layer with and without coating, respectively and the spectral integration is taken over the wavelength range from 300 nm to 800 nm. The results of the calculations are shown in the Fig. 6(c) for structures with different parameters. One can see that the integral absorption enhancement $G_I$ has similar tendency to transmittance enhancement depicted in Fig. 4(b). Particularly, the absorption efficiency of Ag-NPC grows up with increasing nanoparticle sizes and reaches 30%. The absorption enhancement $G_I$ with Si-NPC is in agreement with the transmittance enhancement $F_I$ and has maxima of approximately 8% for nanoparticles of 30-40 nm radius. In contrast to the integral transmission enhancement shown in Fig.4, we do not see any absorbance suppression for nanoparticles with radius 60-80 nm (see Fig. 6(c)) and absorption enhancement curve does not follow transmission enhancement for these radiuses. This difference can be attributed to the role of local fields in the absorption mechanism: they do not contribute into transmission spectrum but can give significant contribution into the absorption. Although the calculations show positive enhancement of absorption for Si- and Ag-NPCs in the considered geometries, our additional calculations (not shown here) predict that one layer of 50-nm ITO coating on top of the bare photovoltaic structure can increase the absorbance on 43%. Nevertheless, the resonant absorption enhancement at the wavelength $\lambda_p$ can give significantly higher values. For that we introduced the following magnitude: $G_p = \left(\eta(\lambda_p)/\eta_0(\lambda_p) - 1\right) \cdot 100\%$, which is proportional to the conversion enhancement calculated at the $\lambda_p$. Simulation results show the monotonic increase of $G_p$ with increasing of nanoparticle radius reaching the value of 65% (Fig. 6(c)). Thus, one can see that placing NPC on top of active layer of silicon photovoltaic element can efficiently increase absorbance inside it: up to 8-10% with Ag- or Si-NPC of small radius (30-40 nm) and up to 30% with Ag-NPC of radius 80 nm having advantages of multipole resonances of plasmonic localized modes.

## Discussions

We have theoretically studied silver and silicon nanoparticle arrays as antireflective coatings for high-index silicon substrate and clarified the role of induced and intrinsic magnetic moments in the nanoparticle array. For the silicon coatings, under illumination of visible light, electric and magnetic moments are induced in the nanoparticles. Tuning the nanoparticle sizes one can control the position of the transmission maxima owing to strong dependence of dipole Mie resonances on nanoparticle size.

The observed antireflectance in silicon nanoparticle array results from destructive interference of the wave reflected from the substrate surface and light scattered by magnetic and electric dipole modes. One can consider this feature as a substrate-mediated Kerker effect, which manifest itself in zero reflection from the substrate coated with nanoparticle arrays rather than suppressed backward scattering for nanoparticles in vacuum. The reflection suppression is shown for the wavelength range between the magnetic and electric dipole resonances accompanied by resonant increase of the transmission to the substrate. In the case of silver coatings, magnetic moment originates from strong reflection from the substrate, which induces "hot spot" with increased both electric and magnetic fields, referred as the substrate-induced bi-anisotropy. Similar to silicon nanoparticle array, scattering from electric and magnetic moments cancel each other resulting in zero reflection. Overall, the most important properties of silicon-based nanoparticle coatings is the tunability of their spectra caused by a strong dependence of magnetic and electric resonance spectral positions on nanoparticle size, which is not the case for plasmonic nanoparticles. One can speculate that the observed resonant enhancement of transmission with narrow spectral width (in a range from 20 to 70 nm) can be applied in ultra-compact photodetectors and filtering systems. In a more broad perspective, the studied nanoparticle arrays can be utilized in metasurfaces as a functional element in optoelectronic and photovoltaic devices.

**Acknowledgements** The authors are thankful to Alexander Krasnok for sharing the ideas about the work, and to Constantin Simovskii and Yuri Kivshar for fruitful discussions.

**Author Contributions** P.A.B. and M.I.P. suggested ideas and guided the study. K.V.B. and V.E.B. performed numerical simulations and result analysis. All the authors contributed to the manuscript preparation.

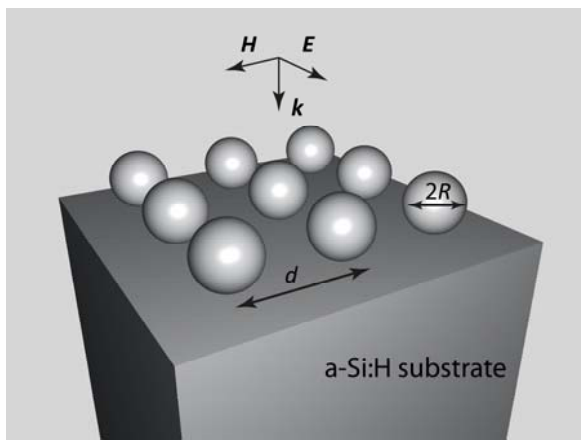

**Figure 1** Square periodic array of spherical nanoparticles with radius $R$ and period $d$ on the top of a-Si:H substrate. The direction of incident wave is shown by the wave vector $\boldsymbol{k}$.

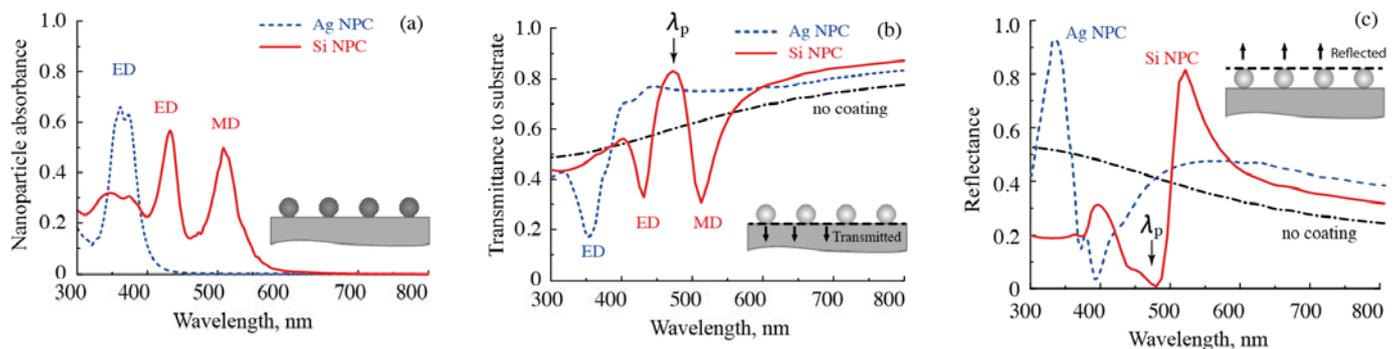

**Figure 2.** (a) Absorbance inside silicon and silver nanoparticles. Inset: nanoparticle array; particles are shaded to show the region of absorbance calculations. (b) Transmittance through the silicon surface without coating (dash-dot black line), or with silicon- (solid red line) and silver- NPC (dashed blue line). Parameters are the same as for (a). The wavelength of resonant transmission for Si-NPC is denoted as $\lambda_p$. Inset: nanoparticle array; dashed line right beneath nanoparticles shows the place of calculation of transmittance in the direction shown by the arrows. (c) Spectral dependences of the reflectance for bare substrate, for silver, and silicon coatings. Inset: nanoparticle array; dashed line right on top of nanoparticles shows where calculations were done. The calculations were performed for Ag-NPC $R_{Ag}$ = 30 nm, $d_{Ag}$ = 125 nm and for Si-NPC $R_{Si}$ = 60 nm, $d_{Si}$ = 250 nm.

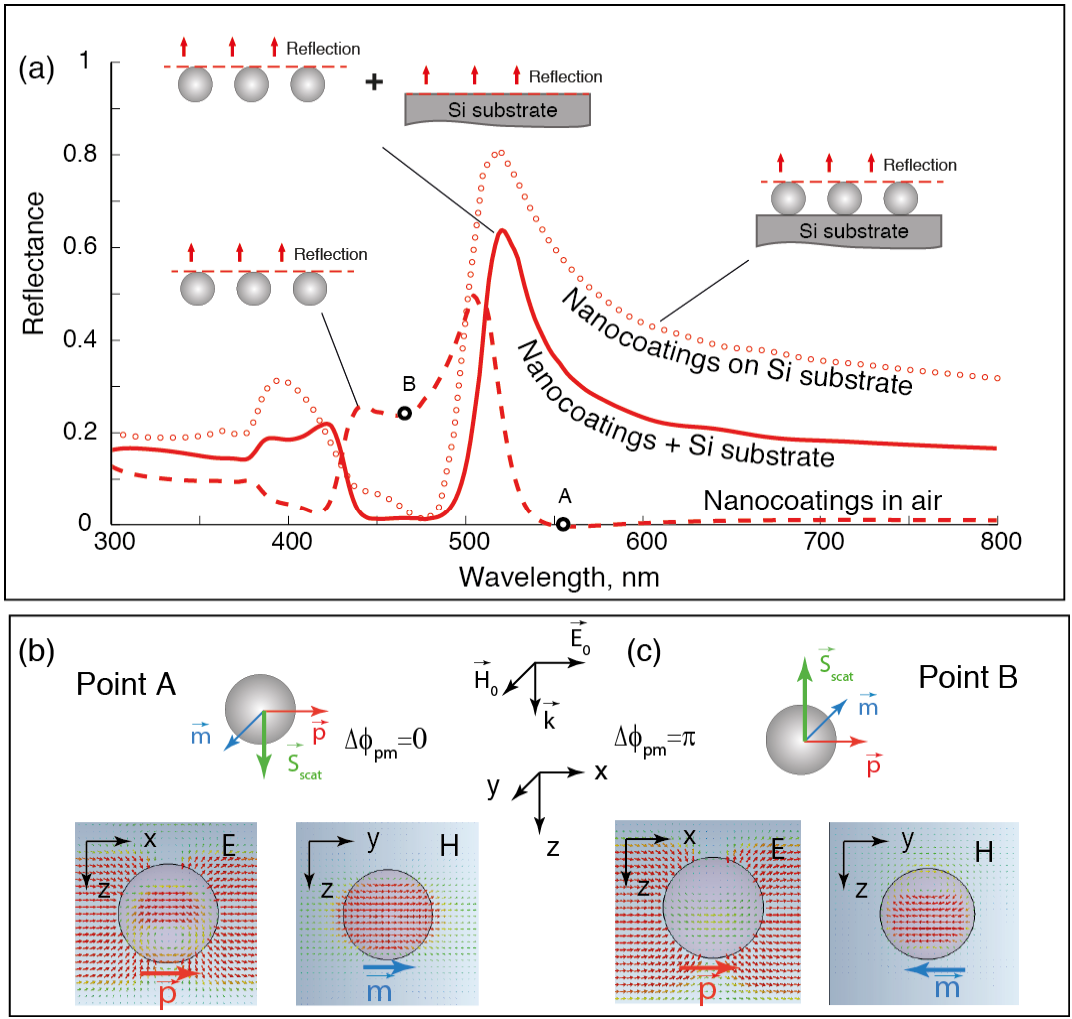

**Figure 3.** (a) Reflectance spectra for a plane wave normally incident: over silicon nanoparticle array in air (dashed line); over silicon nanoparticles on top of the silicon substrate (round circles). The solid line shows the vectorial sum of the fields reflected from the nanoparticle array in air and from the bare silicon substrate. The parameters of the array are identical to the parameters in Fig. 2. The orientation of dipole moments in silicon nanoparticles in air at two wavelengths: (b) corresponding to point A (Kerker effect in vacuum) and (c) to point B (the position of antireflectance of Si-NPC). The corresponding electric and magnetic field distributions are also shown.

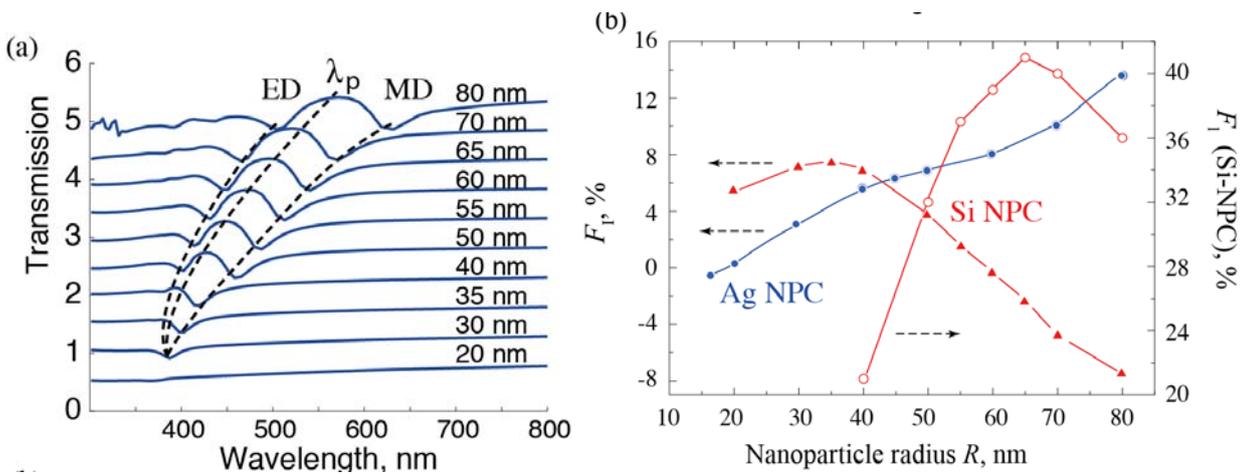

**Figure 4.** (a) The transmittance spectra for Si-NPC with nanoparticles of different radiuses $R$. The positions of ED- and MD-resonances and transmittance enhancement are shown with the black dashed lines. (b) Left axes: integral enhancement $F_I$ for both types of coatings (Eq. (1)). Right axes: the peak enhancement $F_p$ (Eq. (2)) for Si-NPC at $\lambda_P$.

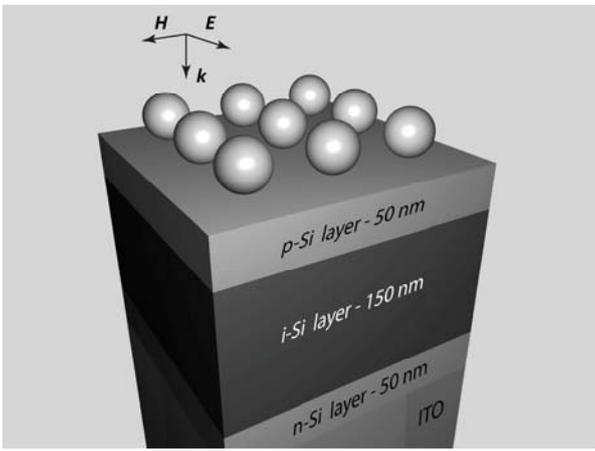

**Figure 5.** The nanoparticle coating on top of thin-film solar cells: active layer consists of 50-nm-thick p-Si, 150-nm-thick i-Si, and 50-nm-thick n-Si layers and it is on top of ITO substrate.

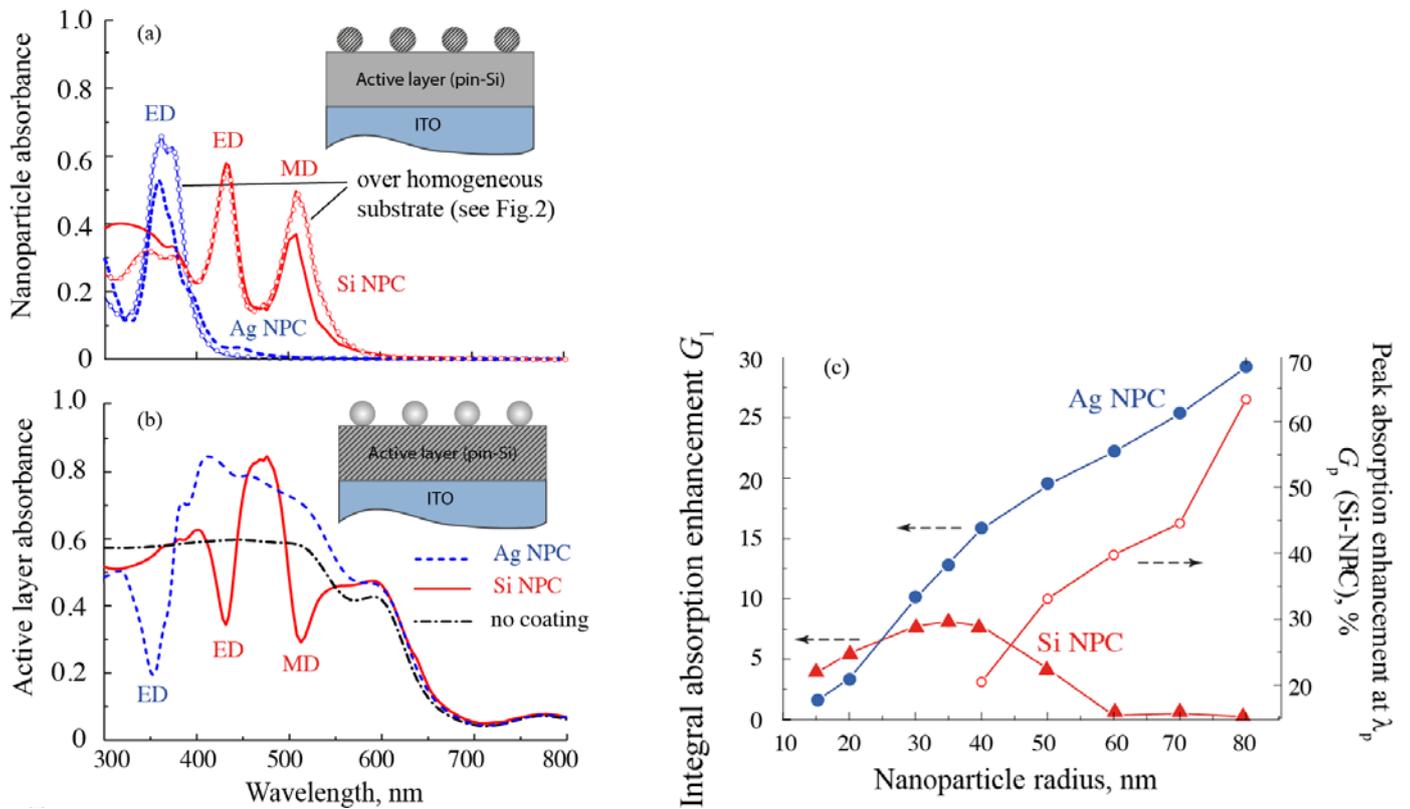

**Figure 6.** (a) The spectra of the light absorption inside the NPC placed over the photovoltaic element shown in Fig. 5. The dashed line corresponds to absorption spectra for NPC placed over the amorphose substrate (also shown in Fig. 2). Inset: schematic view of the structure; shaded region indicates where absorbance is calculated. (b) Absorbance inside the active layer (the region shown shaded in the Inset). (c) The integral absorbance enhancement for Ag-NPC and Si-NPC. The resonant absorbance enhancement at $\lambda_p$ due to Kerker-type effect is also shown at secondary y-axis.